\documentclass[aps,prc,floatfix,twocolumn,superscriptaddress]{revtex4-1}

\usepackage[]{graphicx}
\usepackage{color}

\usepackage{amsmath,amssymb,amsfonts}
\usepackage{hyperref}
\hypersetup{
  colorlinks=true,        
  linkcolor=blue,         
  citecolor=cyan,         
}

\usepackage{slashed}

\begin{document}

\title{Quenching and flow of charm and bottom quarks via semi-leptonic decay of $D$ and $B$ mesons in Pb+Pb collisions at the LHC}

\author{Shu-Qing Li}
\affiliation{School of Physical Science and Intelligent Engineering, Jining University, Qufu, Shandong, 273155, China}

\author{Wen-Jing Xing}
\email{wenjing.xing@mails.ccnu.edu.cn}
\affiliation{Institute of Frontier and Interdisciplinary Science, Shandong University, Qingdao, Shandong 266237, China}

\author{Shanshan Cao}
\affiliation{Institute of Frontier and Interdisciplinary Science, Shandong University, Qingdao, Shandong 266237, China}

\author{Guang-You Qin}
\email{guangyou.qin@ccnu.edu.cn}
\affiliation{Institute of Particle Physics and Key Laboratory of Quark and Lepton Physics (MOE), Central China Normal University, Wuhan, Hubei, 430079, China}

\date{\today}
\begin{abstract}

Heavy flavor particles provide important probes of the microscopic structure and thermodynamic properties of the quark-gluon plasma (QGP) produced in high-energy nucleus-nucleus collisions.
We study the energy loss and flow of charm and bottom quarks inside the QGP via the nuclear modification factor ($R_\mathrm{AA}$) and elliptic flow coefficient ($v_2$) of their decayed leptons in heavy-ion collisions at the LHC.
The dynamical evolution of the QGP is performed using the (3+1)-dimensional viscous hydrodynamics model CLVisc; the evolution of heavy quarks inside the QGP is simulated with our improved Langevin model that takes into account both collisional and radiative energy loss of heavy quarks; the hadronization of heavy quarks is simulated via our hybrid coalescence-fragmentation model; and the semi-leptonic decay of $D$ and $B$ mesons is simulated via PYTHIA.
By using the same spatial diffusion coefficient for charm and bottom quarks, we obtain smaller $R_\mathrm{AA}$ and larger $v_2$ of charm decayed leptons than bottom decayed leptons, indicating stronger energy loss of charm quarks than bottom quarks inside the QGP within our current model setup.

\end{abstract}
\maketitle

\section{Introduction}

A wealth of evidence indicates a color deconfined QCD matter, called quark-gluon plasma (QGP), is produced in high-energy nucleus-nucleus collisions performed at the BNL Relativistic Heavy-Ion Collider and the CERN Large Hadron Collider (LHC)~\cite{Gyulassy:2004zy,STAR:2005gfr,PHENIX:2004vcz,PHOBOS:2004zne,BRAHMS:2004adc,Muller:2012zq}.
Heavy quarks, including charm and bottom quarks, have served as a successful probe of the QGP properties in these energetic collisions~\cite{Dong:2019byy, Andronic:2015wma, He:2022ywp}.
As their masses are much larger than $\Lambda_\mathrm{QCD}$ and the temperature scale of the QGP created at current RHIC and LHC experiments, they are mainly produced from the initial hard collisions between nuclei, whose cross sections can be calculated within the perturbative QCD (pQCD) framework.
After being produced, heavy quarks travel through and interact with the QGP with their flavors conserved before forming heavy flavor hadrons upon exiting the QGP.
At high transverse momentum ($p_\mathrm{T}$), heavy quarks are expected to lose energy inside the QGP via elastic and inelastic scatterings with the medium constituents, similar to energy loss of light quarks and gluons but with mass effect introduced~\cite{Djordjevic:2013pba,Cao:2016gvr,Xing:2019xae}.
At low $p_\mathrm{T}$, heavy quarks are expected to diffuse inside the QGP and evolve towards thermal equilibrium with the QGP~\cite{Moore:2004tg,Cao:2013ita,Xing:2021xwc,Liu:2021dpm}. As a consequence, heavy quarks can inherit significant amount of flow from the expanding medium, which is usually anisotropic in heavy-ion collisions.
At intermediate $p_\mathrm{T}$, the hadronization process is crucial in forming the final heavy flavor hadron observables, including their spectra, collective flow coefficients and chemical components~\cite{Plumari:2017ntm, He:2019vgs, Cho:2019lxb, Cao:2019iqs,Zhao:2023nrz}.
Experimental measurements on heavy flavor mesons and their decay products, such as non-prompt $D$, $J/\psi$ and leptons, have shown significant suppression of their yields~\cite{STAR:2014wif, STAR:2017kkh, CMS:2017qjw, STAR:2018zdy, PHENIX:2022wim} in nucleus-nucleus (A-A) collisions with respect to proton-proton (p-p) collisions. Strong anisotropic flow coefficients have also been observed~\cite{STAR:2017kkh, CMS:2017vhp, ALICE:2017pbx, ALICE:2020iug}.
The combined study of heavy flavor particle yield modification and anisotropic flow coefficient can provide valuable constraints on the interaction strength (diffusion coefficient) and dynamics (contributions from elastic and inelastic processes) between heavy quarks and the QGP, as well as various thermodynamic properties of the QGP (e.g. viscosity and equation of state)~\cite{Gossiaux:2006yu, Qin:2009gw, Das:2010tj, Uphoff:2011ad, He:2011qa, Young:2011ug, Alberico:2011zy, Fochler:2013epa, Nahrgang:2013saa, Cao:2013ita, Djordjevic:2013xoa, Cao:2015hia, Das:2015ana, Song:2015ykw, Cao:2016gvr, Kang:2016ofv, Prado:2016szr, Cao:2017crw, Xu:2017obm, Liu:2017qah, Rapp:2018qla, Cao:2018ews, Li:2018izm, Ke:2018tsh, Li:2019wri, Katz:2019fkc, Li:2020kax, Chen:2021uar, Liu:2021dpm, Yang:2023rgb, Liu:2023rfi}.

The yield suppression of heavy flavor particles can be quantified by the nuclear modification factor,
$R_\mathrm{AA}(p_\mathrm{T}) = \frac{dN_{\rm AA}/dp_\mathrm{T}}{\langle N_{\rm coll}\rangle dN_{\rm pp}/dp_\mathrm{T}}$,
where $\langle N_{\rm coll} \rangle$ is the average number of the binary nucleon-nucleon collisions per nucleus-nucleus collision for a given centrality class, which is usually calculated using the Glauber model~\cite{Miller:2007ri}.
If there is neither cold nor hot nuclear matter effects, $R_\mathrm{AA}$ should be unity, which has been verified by measurements of vector boson productions at RHIC and the LHC.
Contrarily, clear suppression has been observed for the yield of high $p_\mathrm{T}$ hadrons and jets in relativistic heavy-ion collisions, which is stronger in more central collisions~\cite{PHENIX:2001hpc, STAR:2003fka, ATLAS:2010isq, ALICE:2010yje, CMS:2012aa}. This can be well understood by the energy loss experienced by hard quarks and gluons inside the QGP, namely jet quenching~\cite{Wang:1991xy, Gyulassy:2003mc, Majumder:2010qh, Qin:2015srf, Blaizot:2015lma, Cao:2020wlm, Cao:2022odi}.
For the charm sector, the measurements of prompt $D$ mesons show suppression similar to light flavor hadrons~\cite{STAR:2014wif, CMS:2017qjw, STAR:2018zdy}; while for the bottom sector, the measurements of $B$ mesons, $B$-decayed $D$ and $J/\psi$ also show substantial suppression, though less suppression than prompt $D$ mesons at not very high $p_\mathrm{T}$ due to the mass effect in heavy quark energy loss~\cite{CMS:2018bwt, ALICE:2022tji, ALICE:2023gjj}.

The anisotropic flow of particles produced in high-energy nuclear collisions is quantified by performing Fourier expansion of the azimuthal angle distribution of particles as ${dN}/{d\phi} \propto 1 + 2 \sum_n v_n \cos[n(\phi - \Phi_n)]$,
where $v_n$ is $n$-th order anisotropic flow coefficient and $\Phi_n$ is the corresponding event plane angle.
The average almond shape of the collision zone in the transverse plane at non-zero impact parameter leads to the elliptic flow coefficient $v_2$. Meanwhile, event-by-event quantum fluctuations of nucleonic and sub-nucleonic degrees of freedom contribute to higher-order odd harmonic components. They are also the origin of anisotropic flow in ultra-central collisions~\cite{Alver:2010gr, Qin:2010pf}.
Heavy-ion experiments at RHIC and the LHC have observed strong anisotropic flows for charged and identified hadrons produced from the bulk matter~\cite{STAR:2000ekf, STAR:2001ksn, PHENIX:2002hqx, ALICE:2010suc, ALICE:2011ab, ATLAS:2011ah}, which have been successfully explained by relativistic hydrodynamics models~\cite{Romatschke:2017ejr, Rischke:1995ir, Heinz:2013th, Gale:2013da, Huovinen:2013wma}. The small specific viscosity extracted from these models suggests the QCD matter produced in relativistic heavy-ion collisions is a strongly-coupled QGP that behaves like a perfect liquid. The elliptic flow of heavy flavor hadrons depends on the combination of several factors.
At high $p_\mathrm{T}$, the heavy flavor flow is mainly driven by the anisotropic energy loss of heavy quarks through different path lengths in different propagation directions.
At low $p_\mathrm{T}$, their flow depends on not only the interaction strength between heavy quarks and the QGP which determines the degree of thermalization of heavy quarks with the medium background, but also the hadronization process that further transfers the medium collectivity to the heavy flavor hadrons.
Experimental data show $D$ mesons and light flavor hadrons share similar $v_2$ at low $p_\mathrm{T}$, indicating the thermal behavior of heavy quarks at low $p_\mathrm{T}$~\cite{STAR:2017kkh, CMS:2017vhp, ALICE:2017pbx, ALICE:2020iug}.

In this work, we focus on heavy flavor leptons, which are predominantly produced from the semi-leptonic decay of heavy flavor hadrons.
In particular, we study the separate contributions from $D$ and $B$ mesons to the suppression and flow of heavy flavor leptons, as recently measured by the ATLAS Collaboration~\cite{ATLAS:2020yxw, ATLAS:2021xtw}. This allows us to study the mass dependence of quark interaction with a color deconfined medium. The rest of this paper is organized as follows.
In Sec.~II, we give a brief introduction to our theoretical framework on heavy flavor production and evolution in heavy-ion collisions.
In Sec.~III, we present our numerical results on the nuclear modification and elliptic flow of heavy flavor electrons and muons produced in Pb+Pb collisions at $\sqrt{s_\mathrm{NN}} = 5.02$~TeV. In the end, we summarize in Sec.~IV.

\begin{figure*}[tb]
\includegraphics[width=0.49\linewidth]{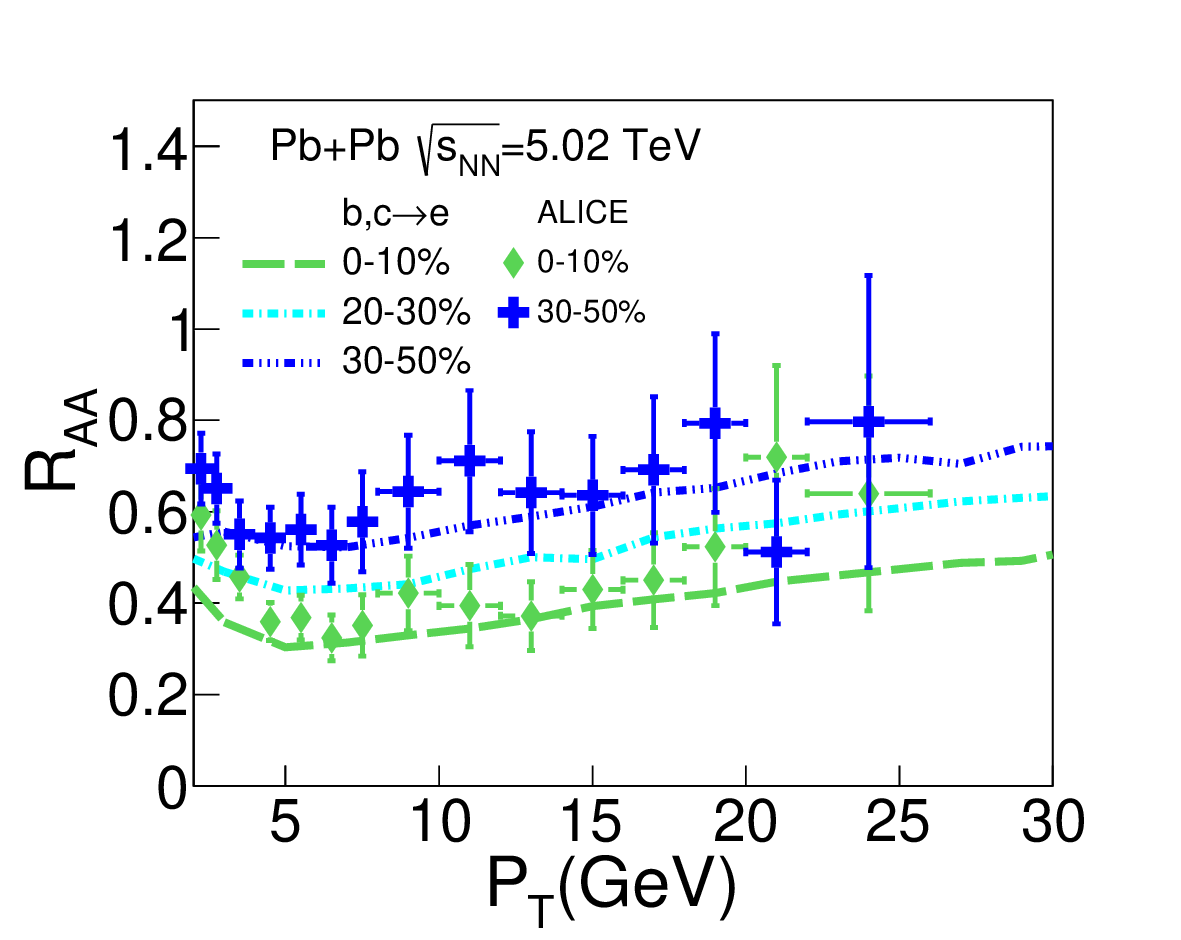}
\includegraphics[width=0.49\linewidth]{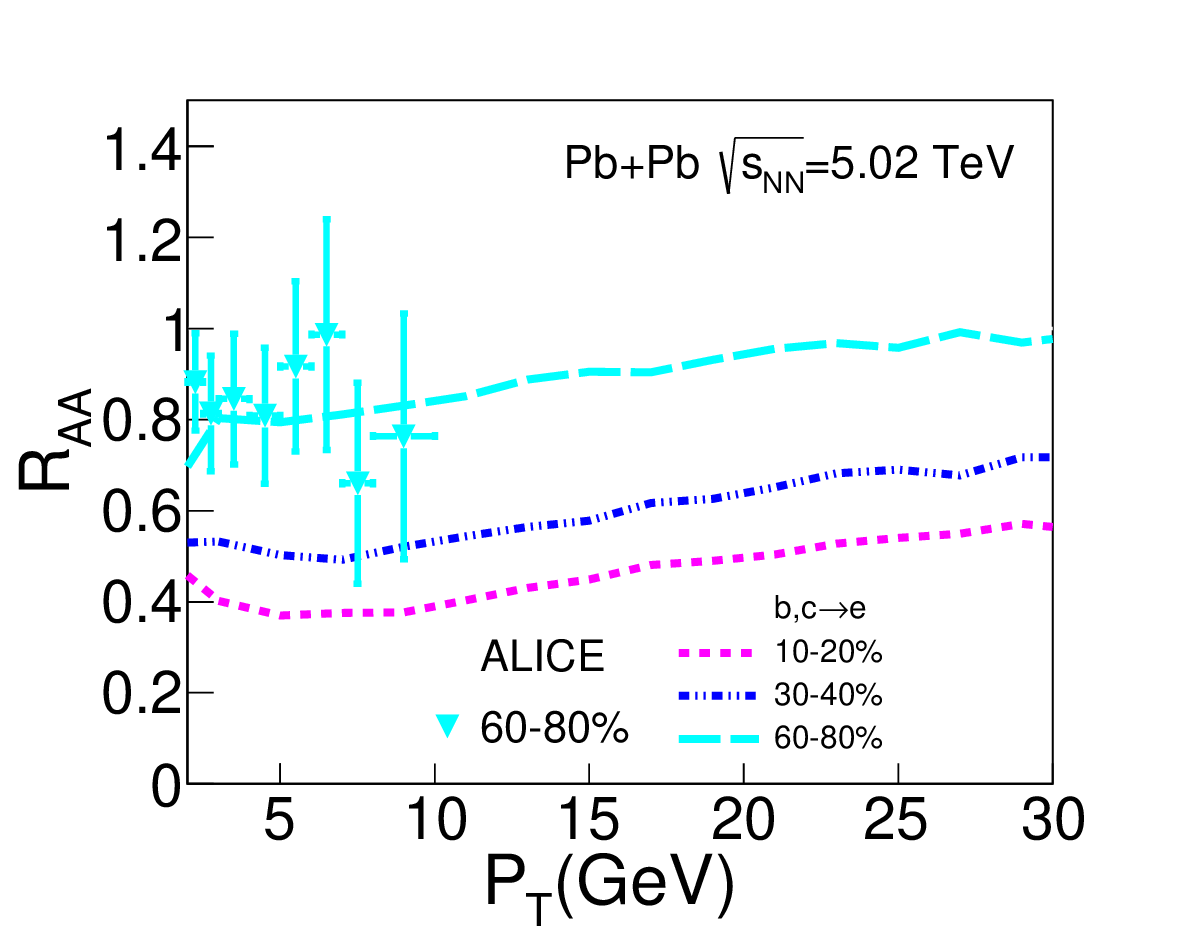}
\includegraphics[width=0.49\linewidth]{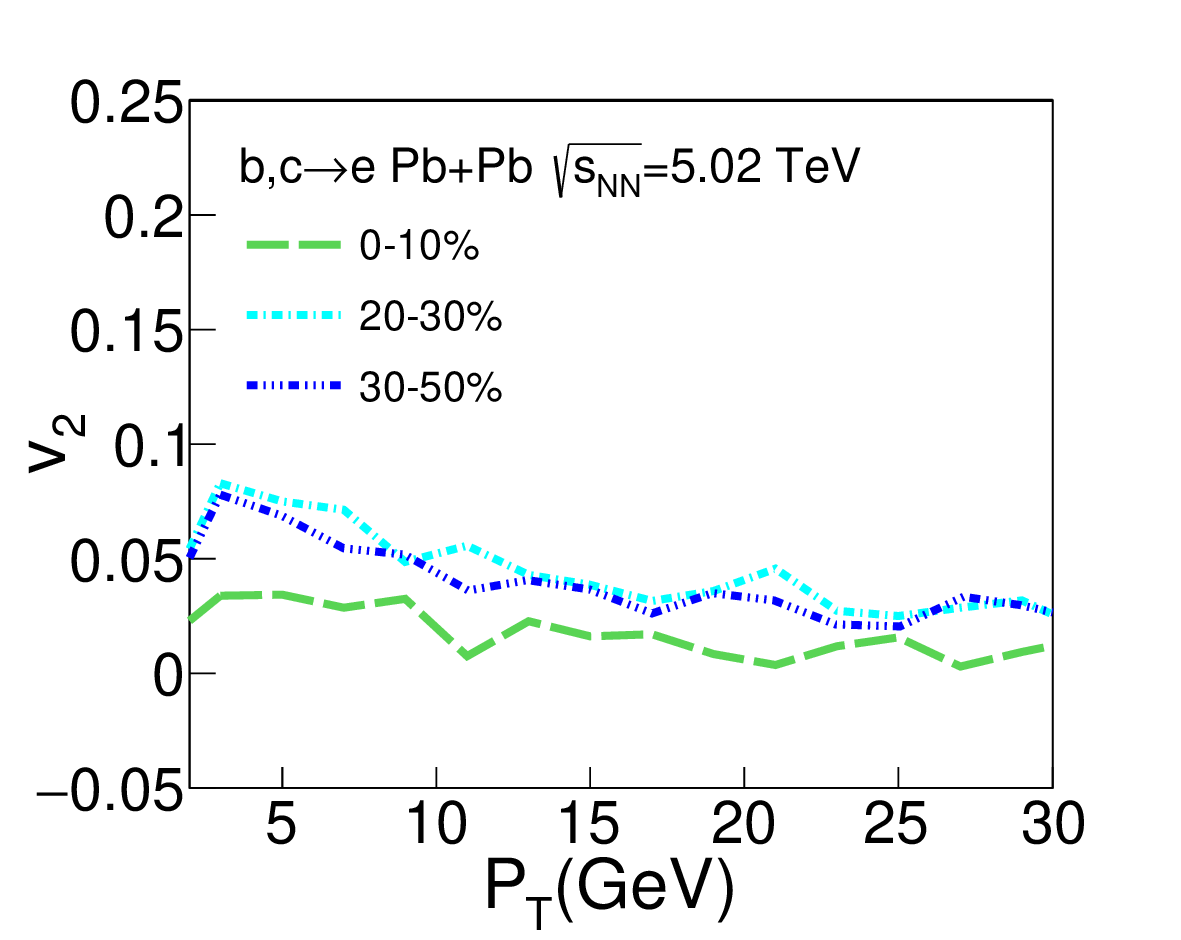}
\includegraphics[width=0.49\linewidth]{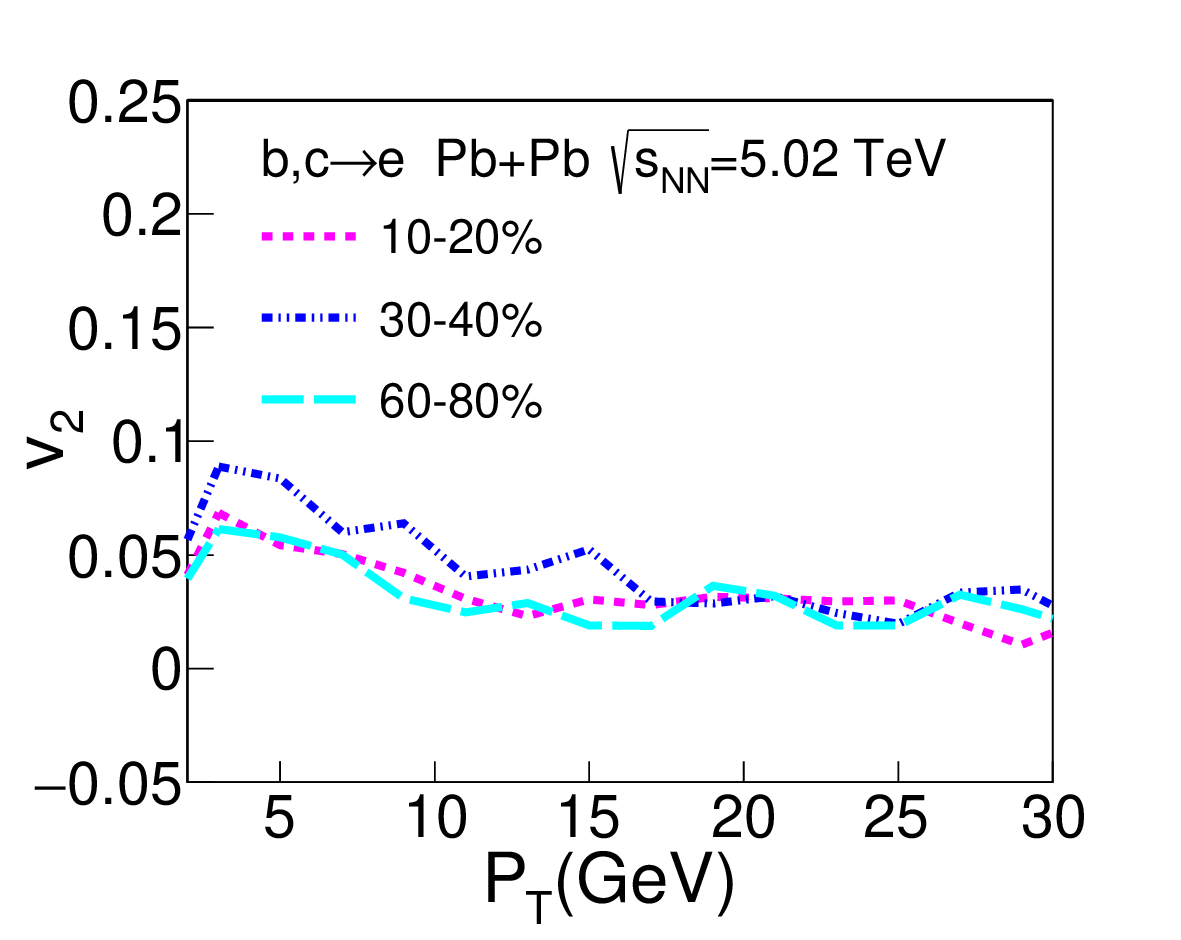}
\caption{(Color online) $R_\mathrm{AA}$ (upper panel) and $v_2$ (upper panel) as functions of $p_\mathrm{T}$ for heavy flavor decayed electrons ($c+b\to e$) in different centrality classes of Pb+Pb collisions at $\sqrt{s_\mathrm{NN}}=5.02$~TeV, compared to the ALICE data~\cite{ALICE:2019nuy}.
} \label{fig1}
\end{figure*}

\begin{figure*}[tb]
\includegraphics[width=0.49\linewidth]{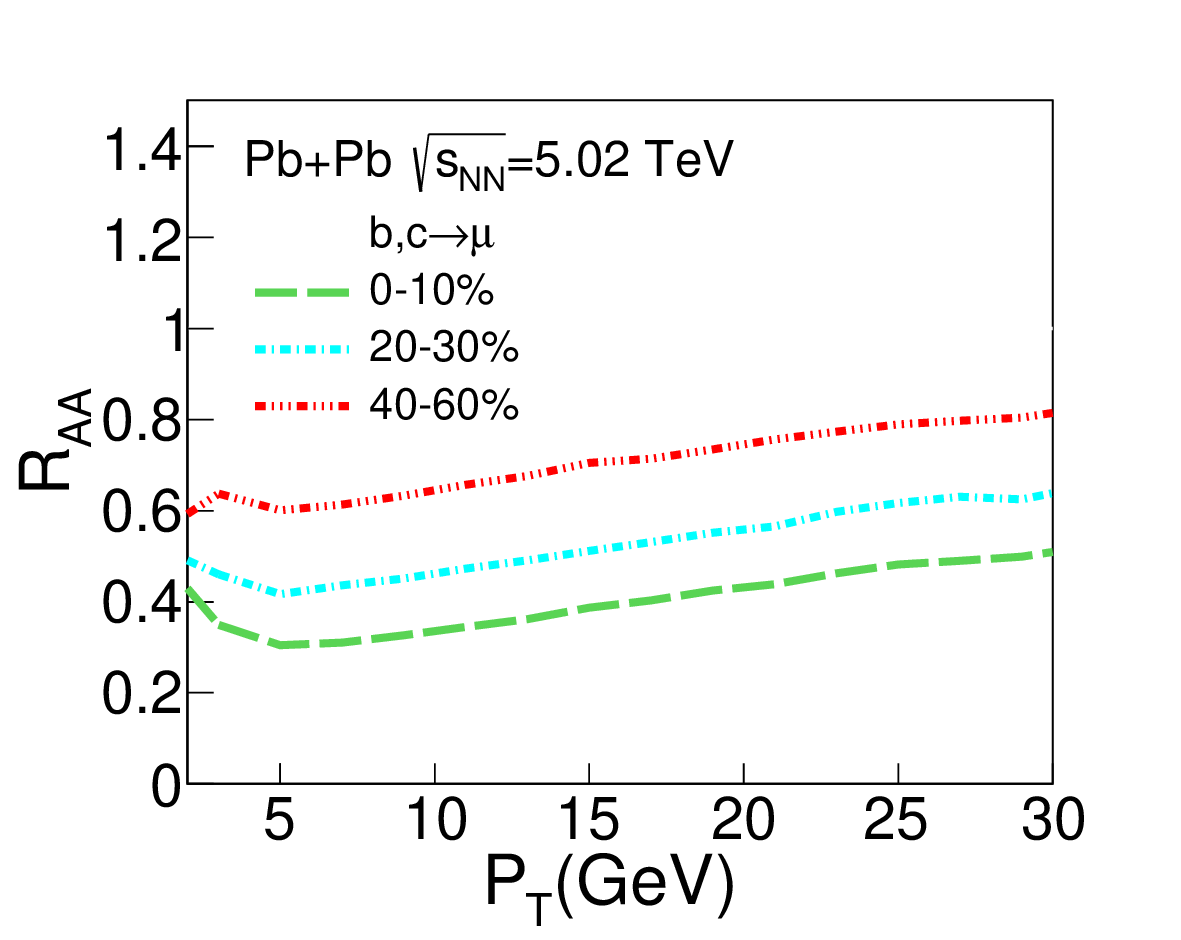}
\includegraphics[width=0.49\linewidth]{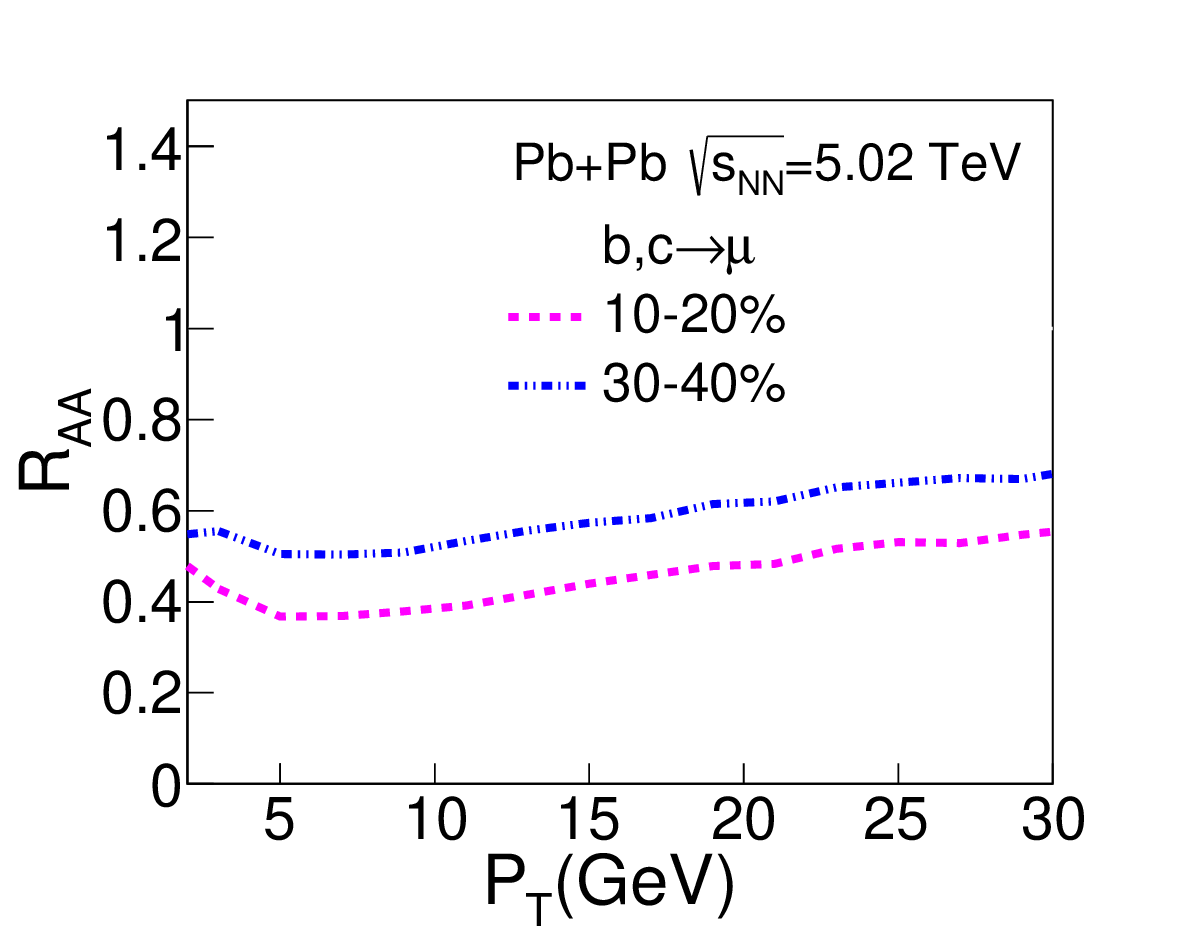}
\includegraphics[width=0.49\linewidth]{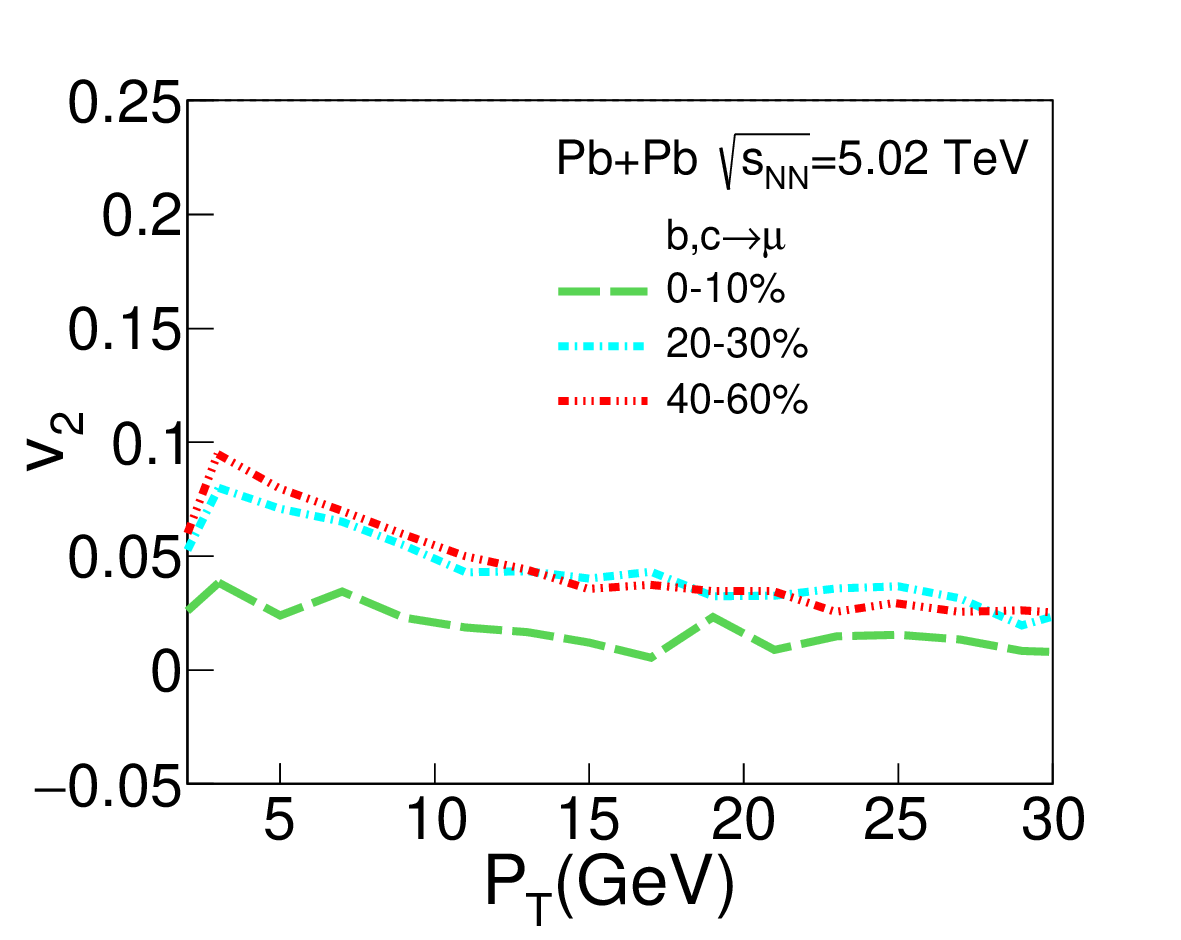}
\includegraphics[width=0.49\linewidth]{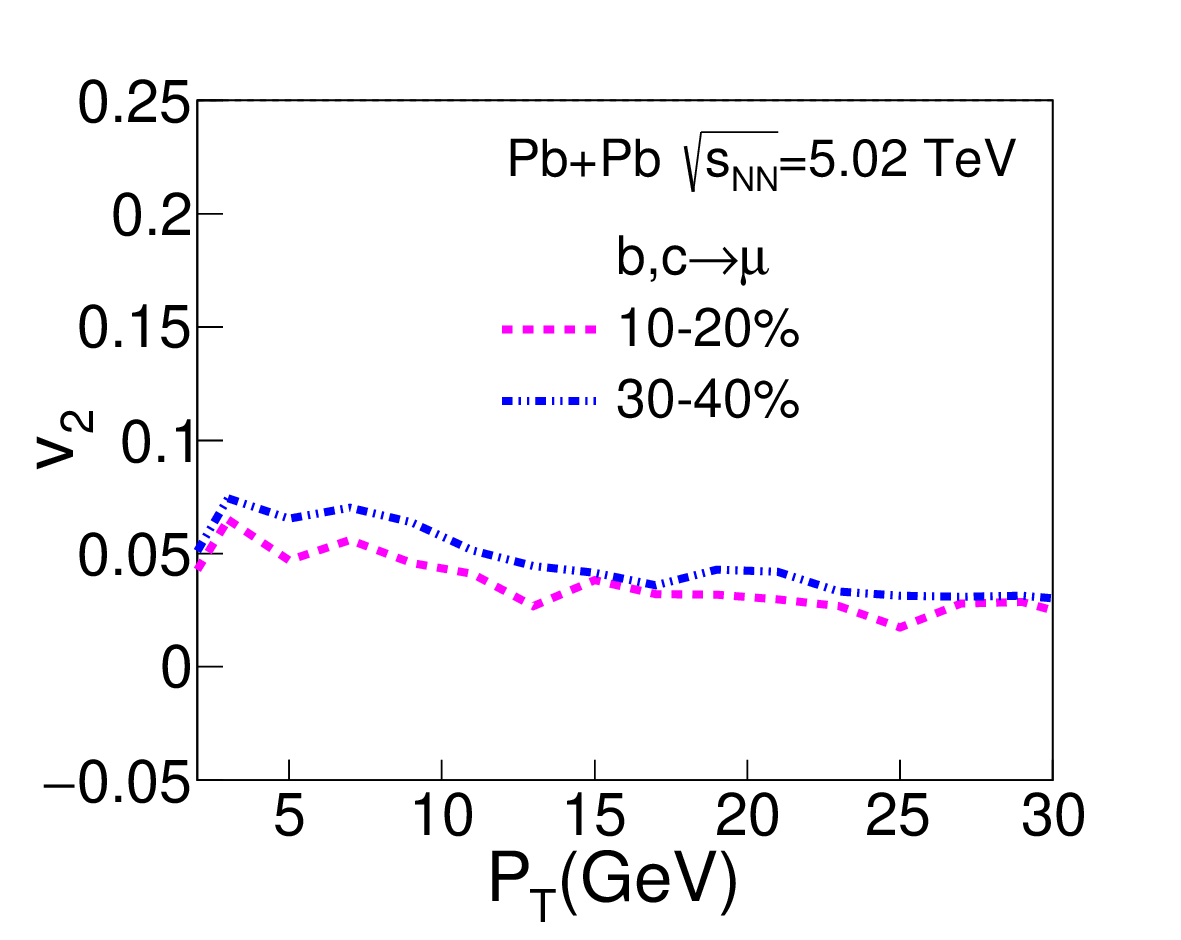}
\caption{(Color online) $R_\mathrm{AA}$ (upper panel) and $v_2$ (lower panel) as functions of $p_\mathrm{T}$ for heavy flavor decayed muons ($c+b\to \mu$) in different centrality classes of Pb+Pb collisions at $\sqrt{s_\mathrm{NN}}=5.02$~TeV.
} \label{fig2}
\end{figure*}

\begin{figure*}[tb]
\includegraphics[width=0.49\linewidth]{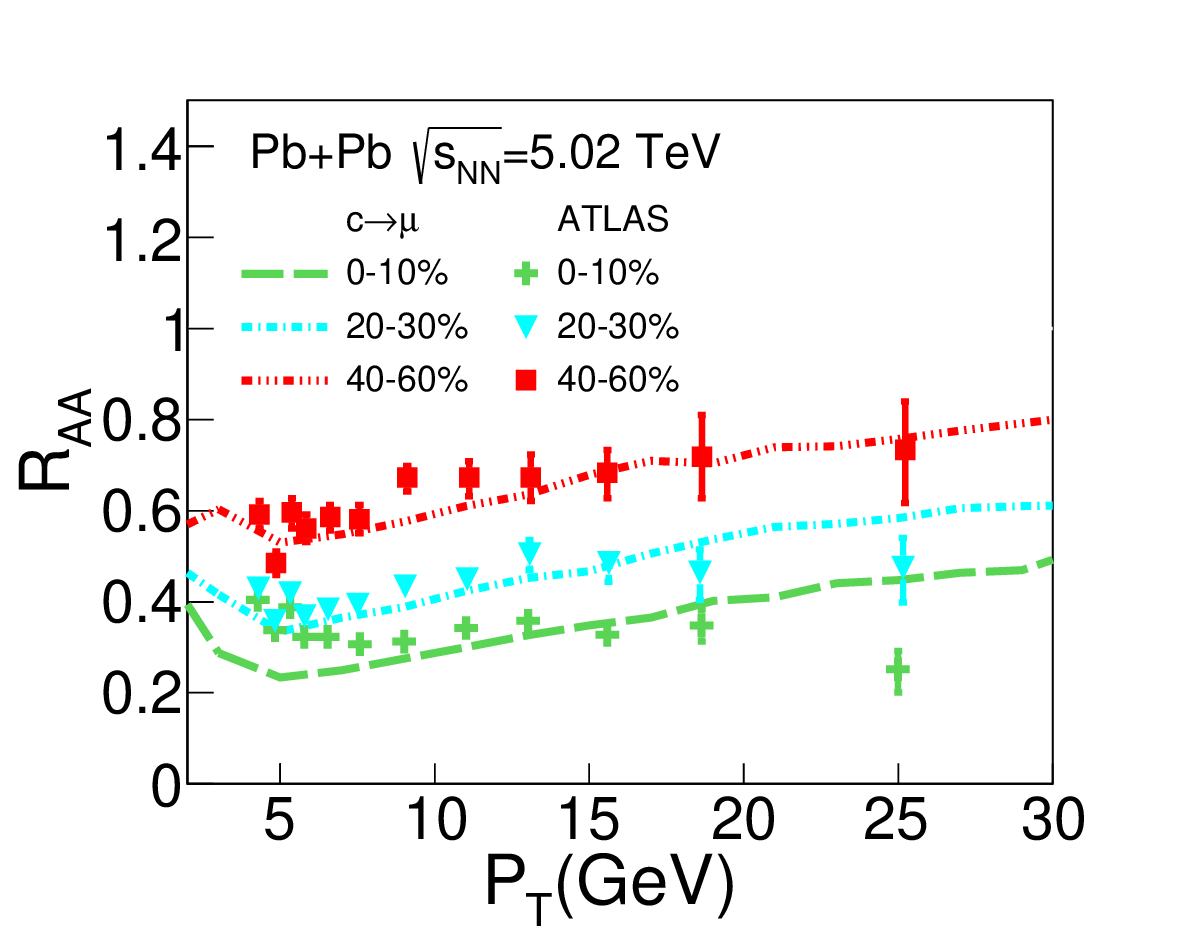}
\includegraphics[width=0.49\linewidth]{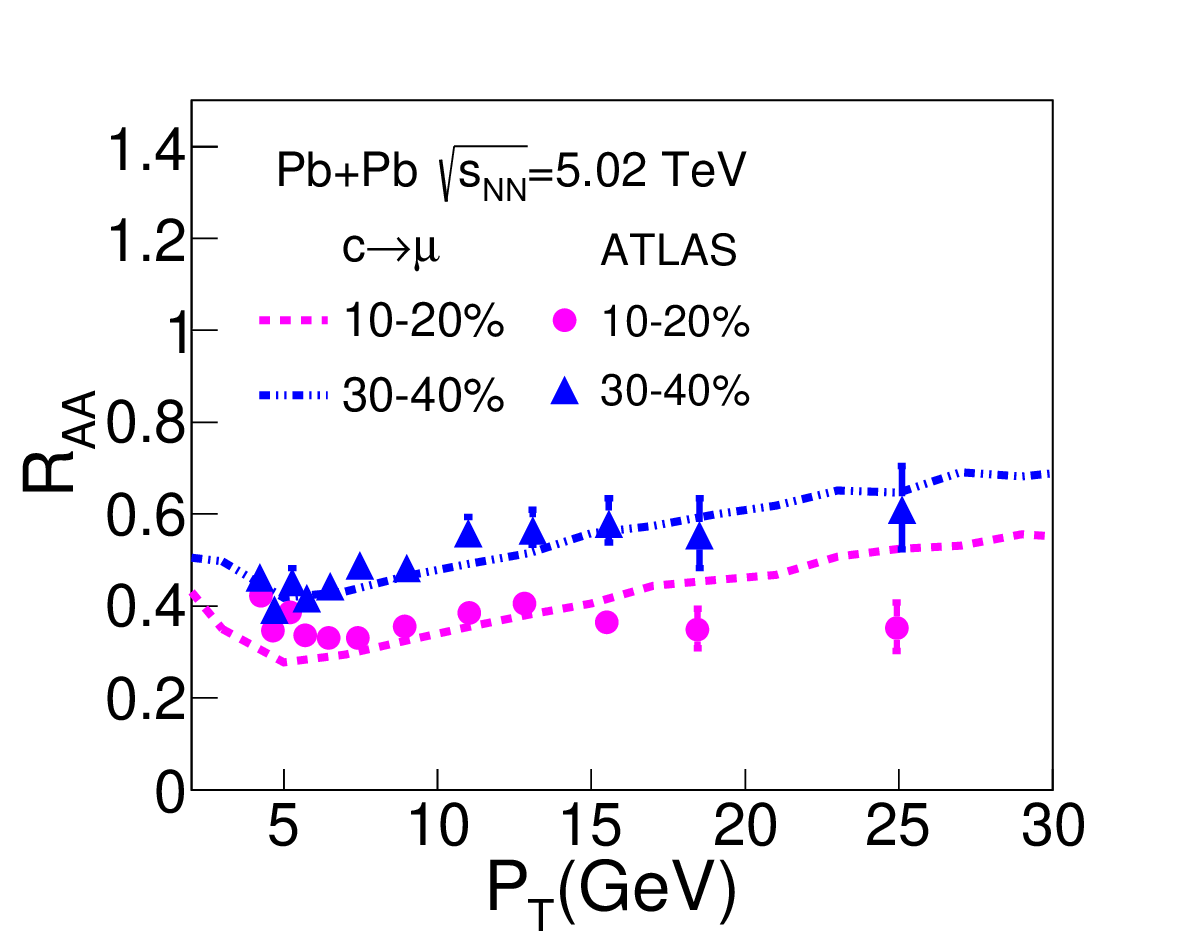}
\includegraphics[width=0.49\linewidth]{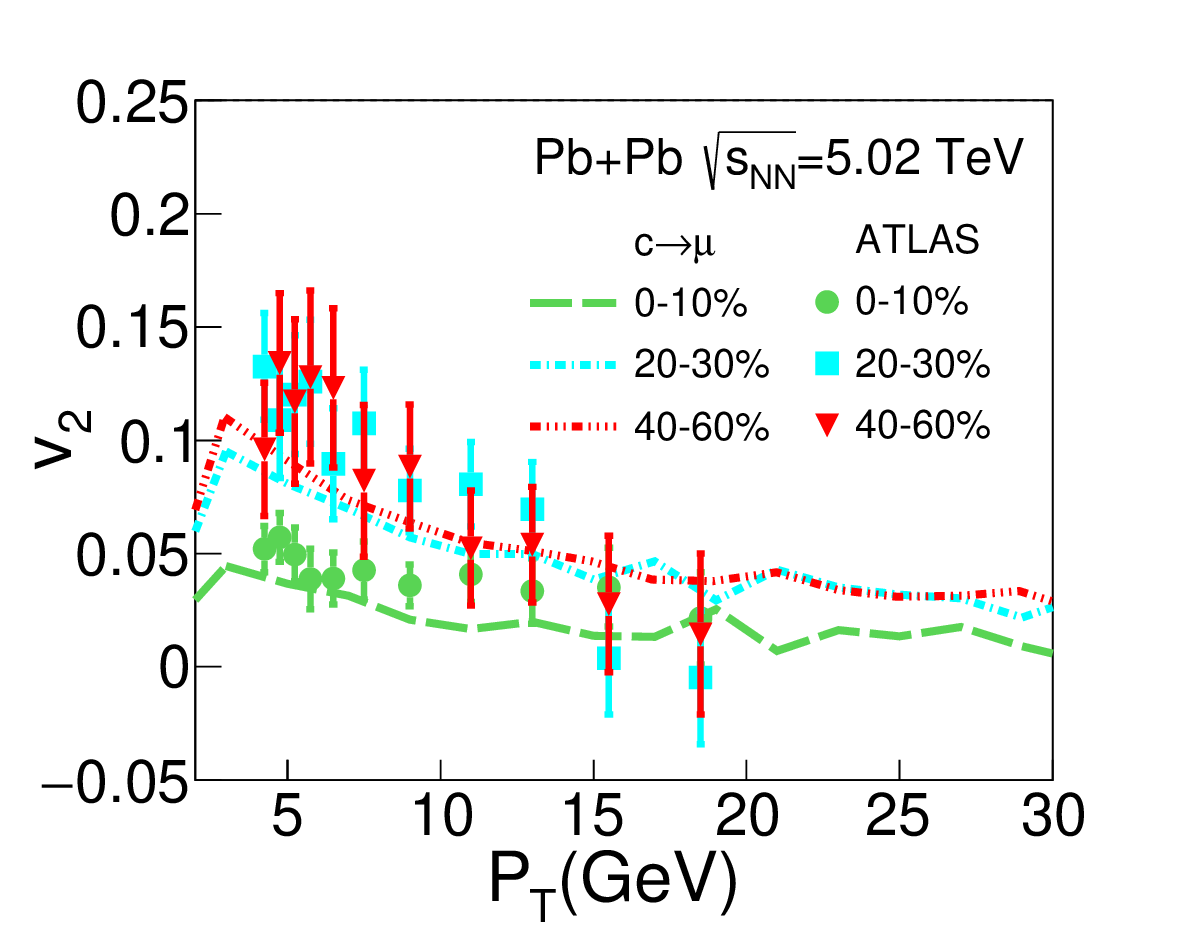}
\includegraphics[width=0.49\linewidth]{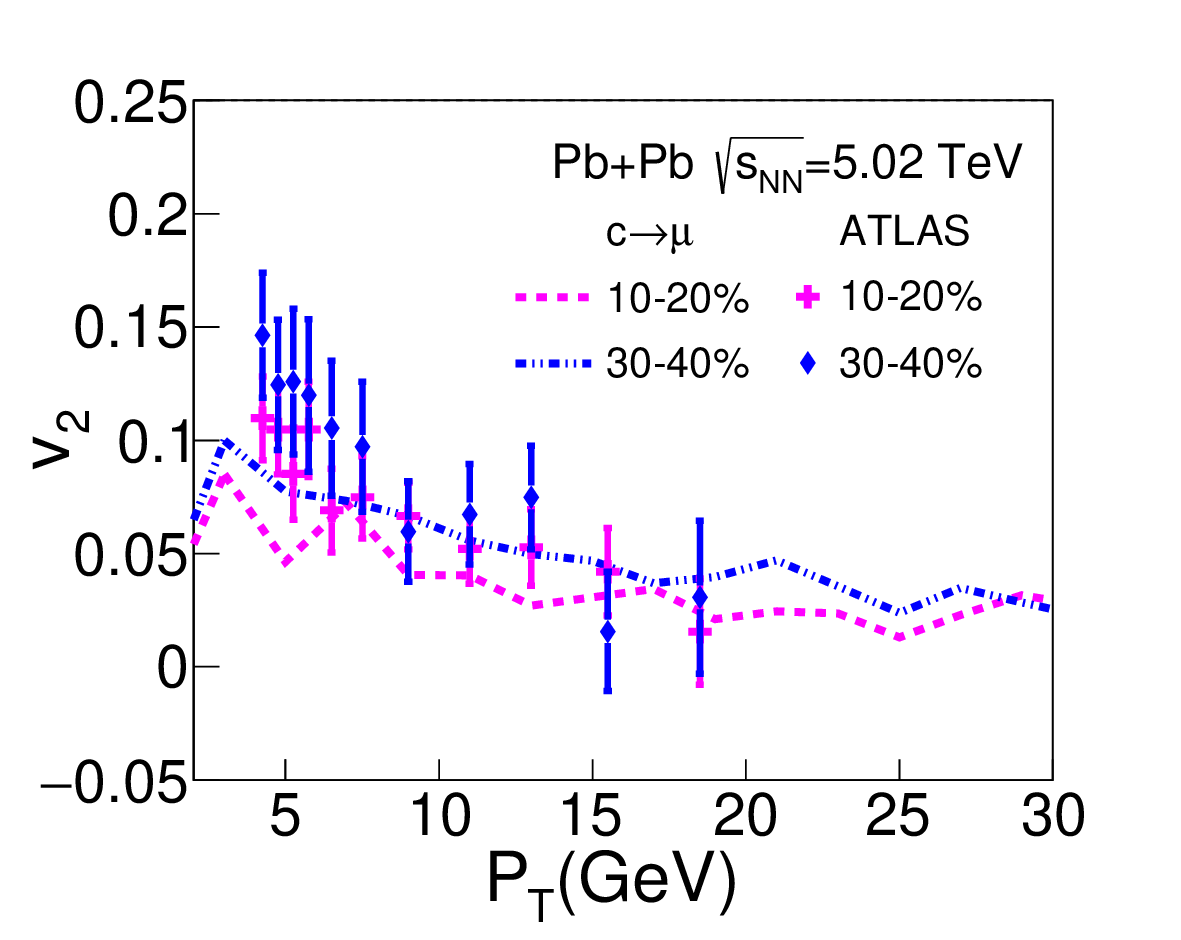}
\caption{(Color online) $R_\mathrm{AA}$ (upper panel) and $v_2$ (lower panel) as functions of $p_\mathrm{T}$ for charm decayed muons ($c \to \mu$) in different centrality classes of Pb+Pb collisions at $\sqrt{s_\mathrm{NN}}=5.02$~TeV, compared to the ATLAS data~\cite{ATLAS:2020yxw, ATLAS:2021xtw}.
} \label{fig3}
\end{figure*}

\begin{figure*}[tb]
\includegraphics[width=0.49\linewidth]{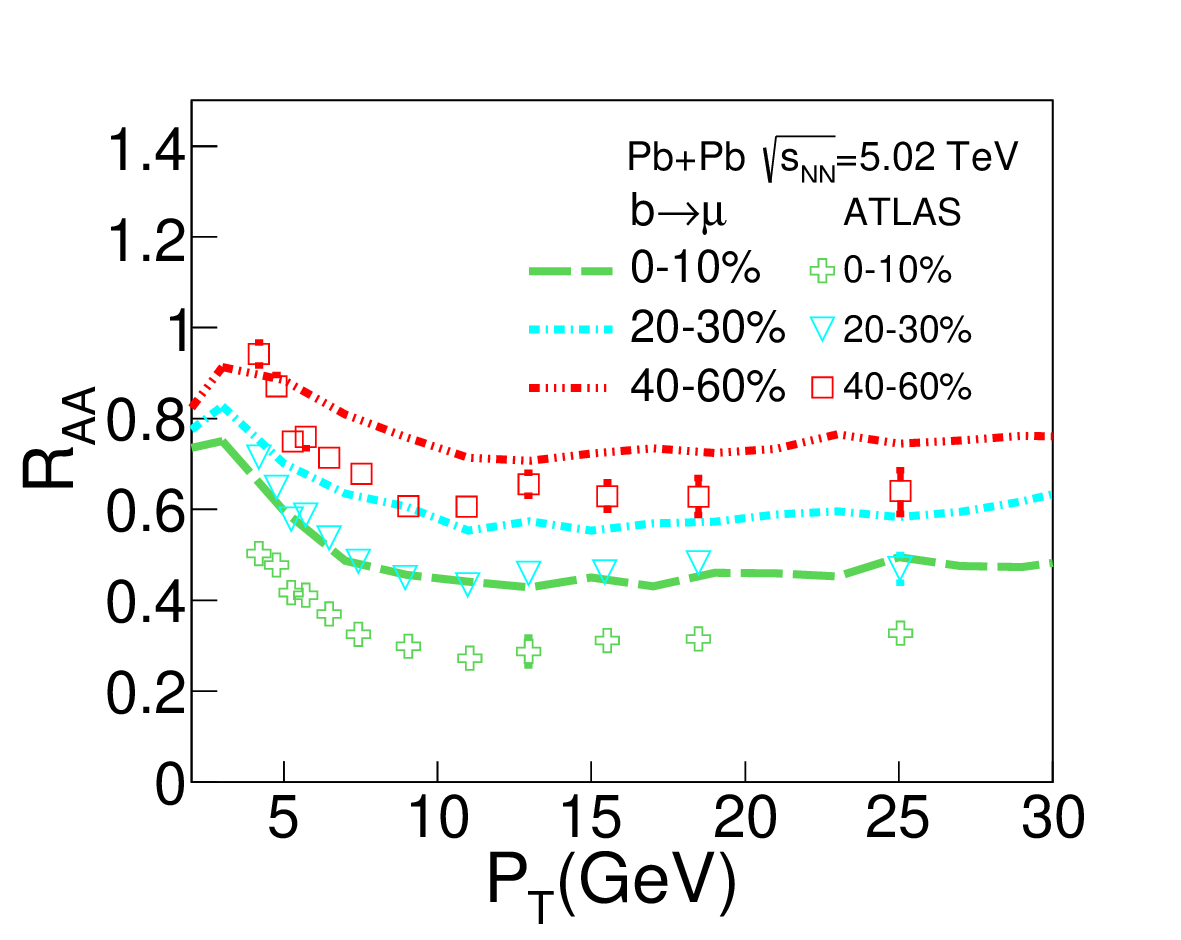}
\includegraphics[width=0.49\linewidth]{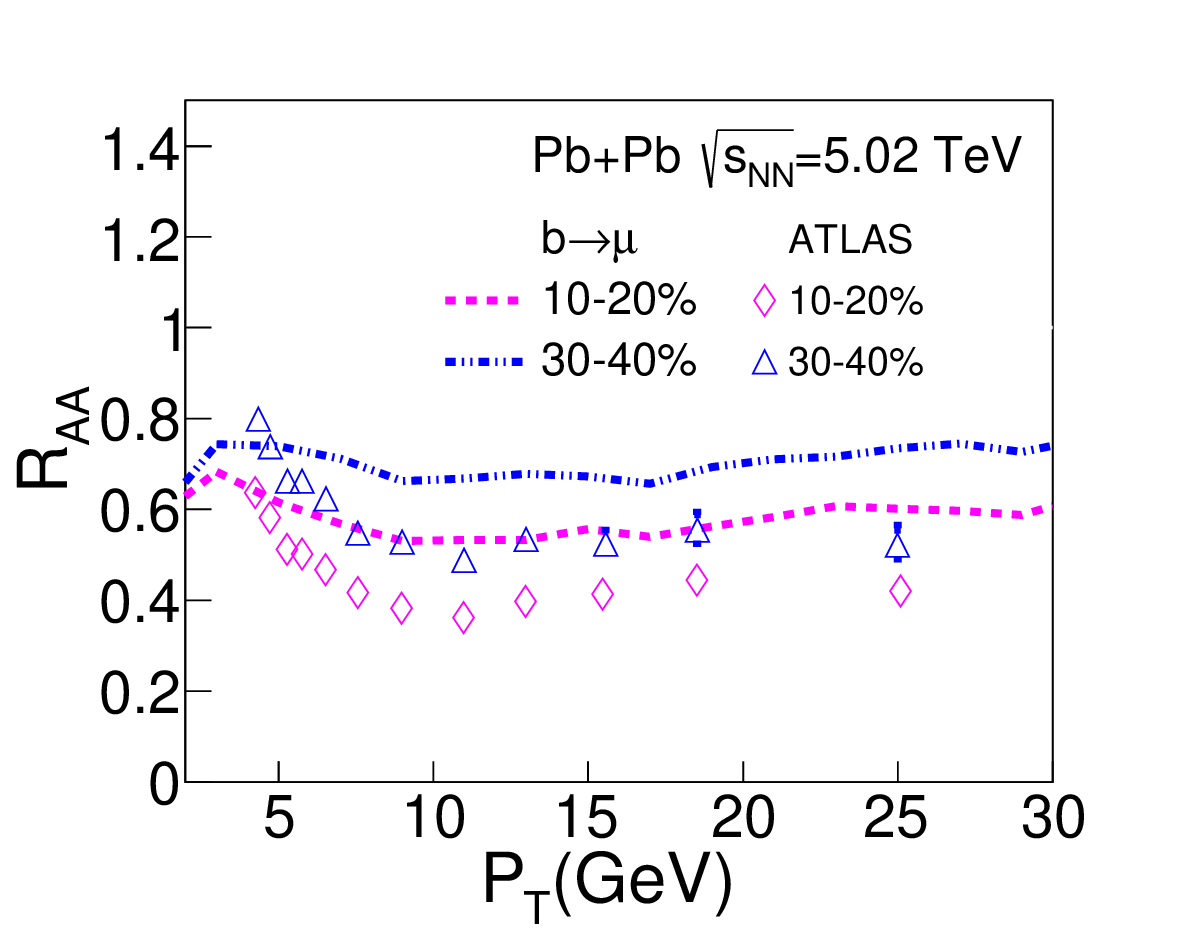}
\includegraphics[width=0.49\linewidth]{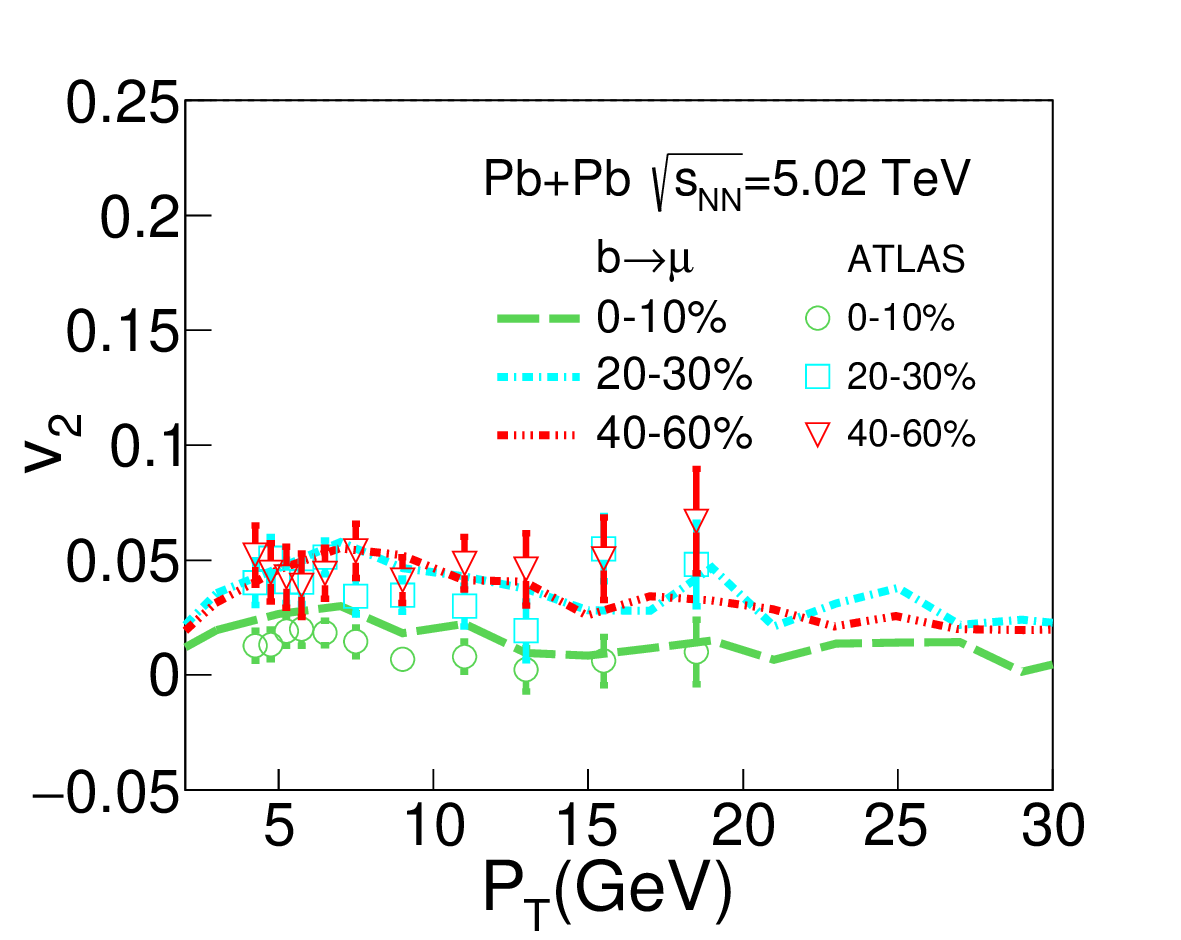}
\includegraphics[width=0.49\linewidth]{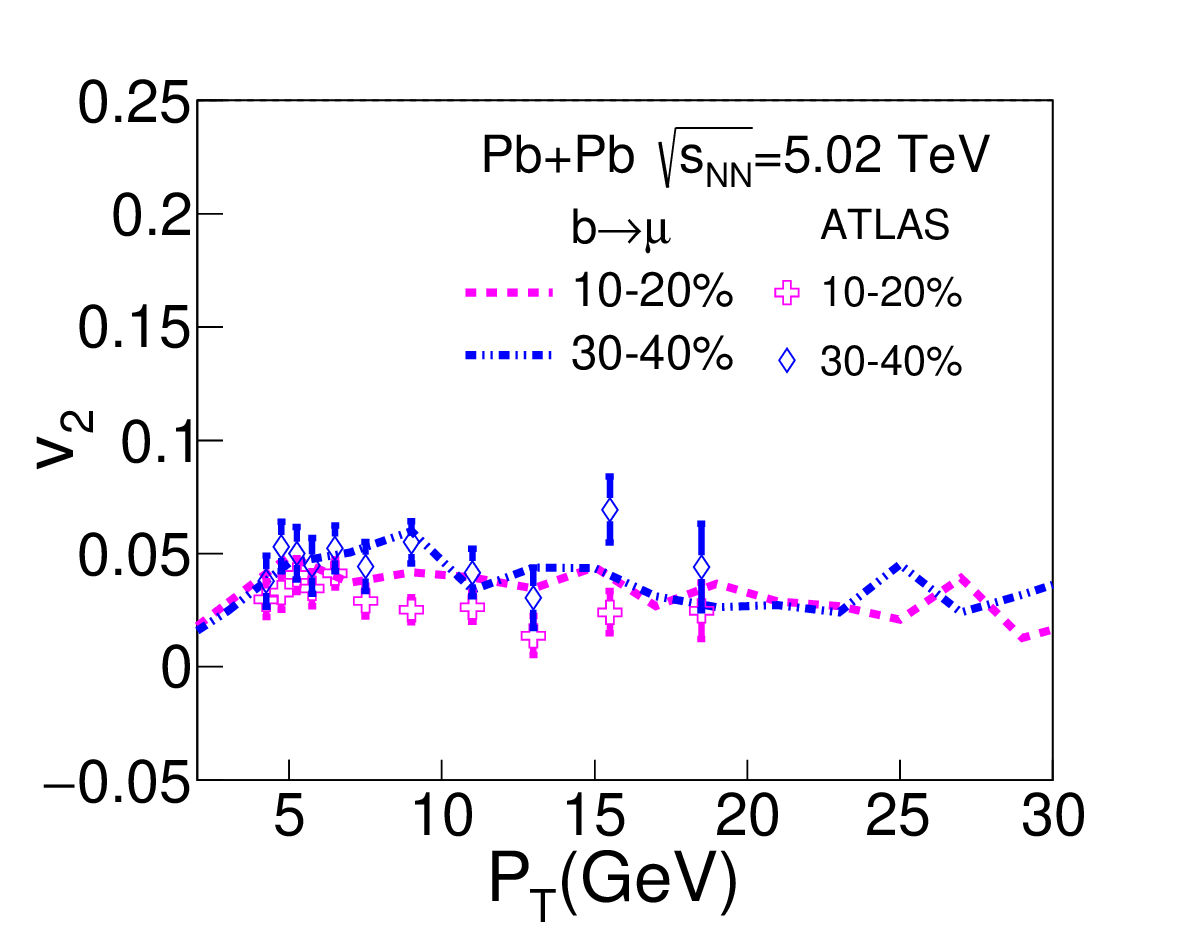}
\caption{(Color online) $R_\mathrm{AA}$ (upper panel) and $v_2$ (lower panel) as functions of $p_\mathrm{T}$ for bottom decayed muons ($b \to \mu$) in different centrality classes of Pb+Pb collisions at $\sqrt{s_\mathrm{NN}}=5.02$~TeV, compared to the ATLAS data~\cite{ATLAS:2020yxw, ATLAS:2021xtw}.
} \label{fig4}
\end{figure*}

\section{Theoretical framework}

In this work, the evolution of heavy quarks inside the QGP is simulated via our improved Langevin approach that solves the following equation~\cite{Cao:2013ita}:
 \begin{align}
  \frac{d\vec{p}}{dt} = -\eta _{D}(p)\vec{p}+\vec{\xi}+\vec{f_{g}}.
 \end{align}
Here, both quasi-elastic scattering and medium-induced gluon bremsstrahlung processes are included.
The first and second terms on the right hand side denote the drag force and thermal random force experienced by heavy quarks, respectively, when they scatter with the QGP, while the third term represents the recoil force $\vec{f}_g = d\vec{p}_g/dt$ exerted on heavy quarks when they emit medium-induced gluons with momentum $\vec{p}_g$.

For the thermal random force $\vec{\xi}$, we assume it to be white noise and satisfy the following correlation:
\begin{align}
\langle\xi^{i}(t)\xi^{j}(t^{\prime})\rangle=\kappa\delta^{ij}\delta(t-t^{\prime}),
\end{align}
where $\kappa$ is the momentum space diffusion coefficient of heavy quarks, characterizing the strength of the thermal force.
It can be related to the drag coefficient $\eta_\mathrm{D}$ via the fluctuation dissipation theorem $\eta_\mathrm{D}(p)=\kappa/(2TE)$. The spatial diffusion coefficient is then given by $D_\mathrm{s} = T/[M\eta_\mathrm{D}(0)]=2T^{2}/\kappa$. By convention, we choose $D_\mathrm{s}$ as the model parameter in our Langevin approach.

For inelastic scattering, the medium-induced gluon radiation probability during a time interval $\Delta t$ can be calculated as:
 \begin{align}
 P_\mathrm{rad}(t,\Delta t) = \langle N_{g}(t,\Delta t)\rangle = \Delta t\int dxdk_{\perp}^{2}\frac{dN_{g}}{dxdk_{\perp}^{2}dt},
 \end{align}
where $x$ and $k_\perp$ are the energy fraction and transverse momentum of the bremsstrahlung gluon with respect to the heavy quark.
Note that in order to interpret the average number of radiated gluons $\langle N_{g}(t,\Delta t)\rangle$ as probability, we choose $\Delta t$ to be sufficiently small such that $\langle N_{g}(t,\Delta t)\rangle <1$. In this study, the medium-induced gluon spectrum is taken
from the higher-twist (HT) energy loss formalism~\cite{Guo:2000nz,Majumder:2009ge,Zhang:2003wk,Zhang:2018nie}:
\begin{align}
\frac{dN_{g}}{dxdk_{\perp}^{2}dt}=\frac{2\alpha_\mathrm{s}P(x)k_\perp^4 \hat q}{\pi ({k_{\perp}^{2}+x^{2}M^{2}})^{4}}\sin^{2}\left(\frac{t-t_{i}}{2\tau _{f}}\right),
\end{align}
where $\alpha_\mathrm{s}$ is the strong coupling strength, $P(x)$ is the parton splitting function, $\hat q$ is the (gluon) jet transport coefficient~\cite{Bass:2008ch,Burke:2013yra,Cao:2021keo} which relates to $\kappa$ via $\hat q = 2\kappa C_{A}/C_{F}$ with $C_A$ and $C_F$ being color factors of gluon and quark, $t_i$ is the initial time of forming the current gluon emission (or the time of the previous emission), and $\tau _{f}=2Ex(1-x)/(k_{\perp}^{2}+x^{2}M^{2})$ is the average formation time of the gluon with $E$ and $M$ being the energy and mass of heavy quarks.
We impose a cutoff for the radiated gluon energy ($E_g = xE > \pi T$) to balance the gluon emission and absorption processes around the thermal scale as dictated by detailed balance.
Note that our improved Langevin approach has one free parameter, the dimensionless parameter $D_\mathrm{s} (2\pi T)$, which characterizes the strength of heavy quark interaction with the QGP.
Other quantities, such as $\kappa$, $\hat{q}$, and $\eta_\mathrm{D}$, can be calculated via their mutual relations.
In this work, we follow our previous studies~\cite{Li:2020kax, Li:2021xbd} to take $D_\mathrm{s}(2\pi T)=4$ for both charm and bottom quarks, which provides a reasonable description of the quenching and elliptic flow of $D$ and $B$ mesons produced in Pb+Pb collisions at the LHC. Note that $D_\mathrm{s}(2\pi T)$ can depend on both the heavy quark energy and the medium temperature, and is not necessarily the same for charm and bottom quarks~\cite{Dang:2023tmb}. This will be left for our future improvement for this Langevin approach.

Before evolving heavy quarks through the QGP, the initial momentum distribution of heavy quarks is calculated using the fixed-order-next-to-leading-log (FONLL) code~\cite{Cacciari:2001td, Cacciari:2012ny, Cacciari:2015fta}, with parton distribution functions taken from CT14NLO~\cite{Dulat:2015mca}.
The initial spatial distribution of heavy quarks is taken from the binary collision vertices determined using the Monte-Carlo Glauber model~\cite{Miller:2007ri}.
The dynamical evolution of the QGP is simulated via the (3+1)-dimensional viscous hydrodynamics code CLVisc~\cite{Pang:2009zm, Pang:2018zzo, Wu:2021fjf}, where the initial condition is calculated using the TRENTo model~\cite{Moreland:2014oya}.
In hydrodynamics simulation, the specific shear viscosity is set as $\eta_\mathrm{v}/s = 0.16$, the relaxation time is set as $\tau_{\pi} = {3\eta_\mathrm{v}}/{(sT)}$, and the equation of state is taken from the {\rm s95-pce-165} parametrization~\cite{Huovinen:2009yb}.
Heavy quarks start interacting with the QGP at the initial time of hydrodynamics evolution ($\tau_0 = 0.6$~fm/$c$) and the interaction stops when the local temperature of the medium drops below $T_\mathrm{c} = 160$~MeV. Upon existing the QGP medium, heavy quarks are converted to heavy flavor hadrons using a hybrid model that takes into account both fragmentation and coalescence processes~\cite{Cao:2019iqs}. In the end, the decay of heavy mesons into leptons is simulated using Pythia~\cite{Sjostrand:2006za}.

\section{Numerical results}

In this section, we present our numerical results on the nuclear modification factor $R_\mathrm{AA}$ and elliptic flow coefficient $v_2$ of heavy flavor electrons and muons produced from semi-leptonic decays of $D$ and $B$ mesons as functions of transverse momentum $p_\mathrm{T}$ in different centrality classes of Pb+Pb collisions at $\sqrt{s_\mathrm{NN}}=5.02$~TeV.

First, we show in Fig.~\ref{fig1} the $R_\mathrm{AA}$ and $v_2$ for heavy flavor decayed electrons ($c+b \to e$) as functions of $p_\mathrm{T}$ in various centrality classes of Pb+Pb collisions at $\sqrt{s_\mathrm{NN}}=5.02$~TeV. For a better visual display, we present 0-10\%, 20-30\% and 30-50\% centralities in the left panels, while 10-20\%, 30-40\% and 60-80\% in the right panels. In general, our model calculation provides a good description of the existing $R_\mathrm{AA}$ data from the ALICE Collaboration.
A strong centrality dependence of both $R_\mathrm{AA}$ and $v_2$ of heavy flavor electrons can be observed here.
The quenching of heavy flavor electrons is stronger (i.e., $R_\mathrm{AA}$ is smaller) in central collisions than in peripheral collisions, which can be understood as the medium size effect on the heavy quark energy loss.
As for $v_2$, it first increases from central to mid-central collisions, but then decreases from mid-central to peripheral collisions.
This results from the combined effect of medium anisotropy (eccentricity) and medium size: the former increases from central to peripheral collisions while the latter decreases from central to peripheral collisions.
Our results also show strong $p_\mathrm{T}$ dependence for both $R_\mathrm{AA}$ and $v_2$ of heavy flavor electrons.
As one moves from low to high $p_\mathrm{T}$, the value of $R_\mathrm{AA}$ typically first decreases and then increases, with its minimum value appearing at intermediate $p_\mathrm{T}$ (around 5~GeV).
In contrast, the value of $v_2$ first increases and then decreases, with its maximum value appearing at intermediate $p_\mathrm{T}$ (around 2 to 3~GeV).
This may be understood with different mechanisms that dominate the heavy flavor spectra at different $p_\mathrm{T}$ regions.
At high $p_\mathrm{T}$, the quenching and elliptic flow of heavy flavor electrons mainly come from the quenching and elliptic flow of $D$ and $B$ mesons at high $p_\mathrm{T}$, which originate from the energy loss of charm and bottom quarks during their interaction with the (anisotropic) QGP.
On the other hand, at low and intermediate $p_\mathrm{T}$, the strong non-perturbative interaction and the coalescence mechanism play more important roles in driving heavy quarks and hadrons towards thermal equilibrium with the medium background. Therefore, the quenching and elliptic flow of heavy flavor decayed electrons are more affected by the radial flow of the QGP medium.

In Fig.~\ref{fig2}, we present similar results as in Fig.~\ref{fig1}, except for heavy flavor decayed muons ($c+b \to \mu$).
Similar to heavy flavor decayed electrons, strong centrality and $p_\mathrm{T}$ dependences are observed for both $R_\mathrm{AA}$ and $v_2$ of heavy flavor decayed muons here.
For the same centrality and $p_\mathrm{T}$, the values of $R_\mathrm{AA}$ and $v_2$ are also similar between heavy flavor decayed electrons and muons, which is expected since the masses of both species of leptons are small compared to those of $D$ and $B$ mesons, and therefore the decay functions to electrons and muons should be similar.

The $R_\mathrm{AA}$ and $v_2$ shown in Figs.~\ref{fig1} and~\ref{fig2} are contributed by the mixture of electrons (muons) decayed from $D$ and $B$ mesons.
In order to separate contributions from charm and bottom quarks, and study the mass effect on heavy quark energy loss, the ATLAS Collaboration has recently measured the $R_\mathrm{AA}$ and $v_2$ of charm decayed and bottom decayed muons separately~\cite{ATLAS:2020yxw, ATLAS:2021xtw}.
Shown in Fig.~\ref{fig3} is our model calculation on the $R_\mathrm{AA}$ and $v_2$ of charm decayed electrons as functions of $p_\mathrm{T}$ in Pb+Pb collisions at $\sqrt{s_\mathrm{NN}} = 5.02$~TeV for various centrality classes, in comparison to the ATLAS data.
Our model reasonably describes the ATLAS data in central, mid-central and peripheral Pb+Pb collisions.
Similar to Fig.~\ref{fig2}, one can observe strong centrality and $p_\mathrm{T}$ dependences of the charm decayed muon $R_\mathrm{AA}$ and $v_2$, as a result of the combined effect of medium size, medium geometry, medium flow and hadronization process.
For the same centrality and $p_\mathrm{T}$, the values of $R_\mathrm{AA}$ here are smaller than those in Fig.~\ref{fig2}, while the values of $v_2$ here are larger than those in Fig.~\ref{fig2}. This is because the contribution from bottom decay is removed here, which yields larger $R_\mathrm{AA}$ and smaller $v_2$ than the contribution from charm decay within our model.

In the end, we present in Fig.~\ref{fig4} the $R_\mathrm{AA}$ and $v_2$ of bottom decayed muons as functions of $p_\mathrm{T}$ in Pb+Pb collisions at $\sqrt{s_\mathrm{NN}} = 5.02$~TeV for different centrality classes, compared to the ATLAS data.
One can see that while our model result reasonably agrees with the $v_2$ data, it overestimates the $R_\mathrm{AA}$, or underestimates the energy loss effect for bottom quarks.
Comparing between Fig.~\ref{fig3} and Fig.~\ref{fig4}, for the same centrality and $p_\mathrm{T}$, we see our model calculation shows larger $R_\mathrm{AA}$ of bottom decayed electrons than charm decayed electrons due to weaker energy loss of heavier quarks within our model, although this hierarchy is not obvious in the current data. This discrepancy may result from the minimal assumption of the constant $D_\mathrm{s}$ parameter we use in the present study. It is now generally acknowledged that $D_\mathrm{s}$ should depend on the medium temperature and the heavy quark momentum, and its value is not necessarily the same between charm and bottom quarks. Including these temperature, momentum and mass dependences of diffusion coefficient should improve our model performance in the future.


\section{Summary}

In this work, we have studied charm and bottom quark energy loss and flow via heavy flavor leptons produced from semi-leptonic decays of $D$ and $B$ mesons.
The (3+1)-dimensional viscous hydrodynamics model CLVisc is used to simulate the dynamical evolution of the QGP produced in Pb+Pb collisions at the LHC.
The evolution of heavy quarks in the QGP is simulated using our improved Langevin model including both collisional and radiative contributions to parton energy loss.
The hadronization of heavy quarks after their escape from the QGP is simulated with our hybrid coalescence-fragmentation model.
The semi-leptonic decay of $D$ and $B$ mesons is simulated using PYTHIA.
Using our heavy quark model, we have calculated the separate contributions from $D$ and $B$ mesons to the suppression and elliptic flow of heavy flavor decayed leptons.
Both our model calculation and the experimental data show strong centrality and $p_\mathrm{T}$ dependences of the $R_\mathrm{AA}$ and $v_2$ of heavy flavor leptons due to the combined effect of system size, medium anisotropy, medium flow and the hadronization process. By using a constant $D_\mathrm{s}(2\pi T)=4$ for both charm and bottom quarks, we find bottom decayed leptons show larger $R_\mathrm{AA}$ and smaller $v_2$ than charm decayed leptons due to weaker energy loss of heavier quarks inside the QGP, though this hierarchy is not apparent in the current ATLAS data. While our model provides a reasonable description of the ALICE data on the mixed heavy flavor decayed electrons and the ATLAS data on the charm decayed muons, it overestimates the $R_\mathrm{AA}$ of the bottom decayed muons observed at ATLAS.
Future more detailed study on heavy flavor hadrons and their decay products, especially with a more sophisticated temperature, momentum and mass dependences of heavy quark diffusion coefficient, can help to better constrain charm versus bottom quark energy loss, the collisional versus radiative contributions to parton energy loss, and transport properties of the QGP in relativistic heavy-ion collisions.


\section*{Acknowledgments}

This work is supported in part by the National Natural Science Foundation of China (NSFC) under Grant Nos. 12225503, 11935007, 11890710, 11890711, 12175122 and 2021-867. W.-J.~X. is supported in part by China Postdoctoral Science Foundation under Grant No. 2023M742099. Some of the calculations were performed in the Nuclear Science Computing Center at Central China Normal University (NSC$^3$), Wuhan, Hubei, China.

\bibliographystyle{plain}
\bibliographystyle{h-physrev5}
\bibliography{refs_GYQ}
\end{document}